\documentclass[aps,prb,twoside,twocolumn,notitlepage,nofootinbib,floatfix,floats,amssymb,amsmath]{revtex4-1}

\usepackage{color}
\usepackage{amsmath}
\usepackage{pifont}  
\usepackage{graphicx}
\usepackage{dcolumn} 
\usepackage{bm}      
\usepackage{amsfonts}
\usepackage{amssymb} 

\begin{document}

\newcommand{\ba}{{\bf a}}
\newcommand{\bbb}{{\bf b}}
\newcommand{\br}{{\bf r}}
\newcommand{\bp}{{\bf p}}
\newcommand{\bq}{{\bf q}}
\newcommand{\bk}{{\bf k}}
\newcommand{\bg}{{\bf g}}
\newcommand{\bs}{{\bf s}}
\newcommand{\bt}{{\bf t}}
\newcommand{\bG}{{\bf G}}
\newcommand{\bP}{{\bf P}}
\newcommand{\bJ}{{\bf J}}
\newcommand{\bK}{{\bf K}}
\newcommand{\bL}{{\bf L}}
\newcommand{\bR}{{\bf R}}
\newcommand{\bS}{{\bf S}}
\newcommand{\bT}{{\bf T}}
\newcommand{\bA}{{\bf A}}
\newcommand{\bH}{{\bf H}}

\newcommand{\ga}{\alpha}
\newcommand{\gm}{\mu}
\newcommand{\gb}{\beta}
\renewcommand{\gg}{\gamma}
\newcommand{\gd}{\delta}
\newcommand{\ep}{\epsilon}
\newcommand{\gl}{\lambda}
\newcommand{\go}{\omega}


\title{Bulk physics entwined with a topological surface state}

\author{N. Klier}
\author{S. Sharma}
\author{O. Pankratov}
\author{S. Shallcross}
\email{sam.shallcross@fau.de}
\affiliation{Lehrstuhl f\"ur Theoretische Festk\"orperphysik, Staudtstr. 7-B2, 91058 Erlangen, Germany.}

\date{\today}

\begin{abstract}

For the IV-VI semiconductor family we derive an exact relation between the microscopic gap edge wave functions of the bulk insulator and the Dirac-Weyl topological surface state wave function, thus obtaining a fully microscopic surface state. We find that the balance of spin-orbit interaction and crystal field in the bulk, and the band bending at the surface, can profoundly influence the surface state spin-momentum locking. As a manifestation of this we predict that the spin texture of the $M$-point Dirac cones of SnTe can be tuned through an unexpectedly rich sequence of spin textures -- warped helical with winding number $\pm 1$, $k_x$ linear, hyperbolic, and $k_y$ linear -- e.g. by tuning the band bending at the surface.

\end{abstract}


\maketitle


\emph{Introduction}: Topological insulators (TI) are a state of matter revealed only by the presence of a surface. As such, these are materials for which the bulk-boundary relationship is of profound importance. A full understanding this relation would involve an explicit route from the microscopic variables of the bulk insulator (spin, angular momentum, parity, etc.) to the emergent effective degrees of freedom of the topological Dirac-Weyl (DW) surface state (pseudospin and chirality). Unfortunately, as most TI's are very complex materials, such a fully explicit connection has escaped analytical analysis. The relationship between microscopic and effective variables in TI's, in particular between spin and pseudospin has, therefore, remained obscure. The purpose of this paper is threefold: (i) to provide an exact relation between bulk and surface wavefunctions for a wide class of TI's; (ii) to show that while the TI energy spectrum is universal (the DW cone) the corresponding wave functions are, due to the bulk-boundary connection, highly non-universal; and (iii) on this basis, and as an example, to demonstrate a hitherto unsuspected richness in the spin texture of the $M$-point Dirac cones on the (111) surface of the topological insulator SnTe\cite{tan12,tan13}.

The IV-VI semiconductor family adopt a simple rocksalt crystal structure and, perhaps uniquely amongst the topological insulators, offer the hope of such a fully microscopic theory. We employ a band structure model proven to describe these materials\cite{vol78,vol83,vol85,pan87,pan90} along with a topological boundary condition\cite{vol85,zha12} leading to an interface equation of the super-symmetric type whose solution yields the DW surface state. In contrast to previous studies\cite{zha12,liu13,zha13} -- in which the bulk electronic structure is also described in terms of effective Dirac Hamiltonian -- our model begins at the tight-binding level and allows us to explicitly express the surface state in terms of the truly microscopic variables: the electron spin and the $p$-orbitals of the constituent atomic species. From this fully microscopic yet analytical approach two interesting observations follow. Firstly, there are two distinct mechanisms that act to entangle spin in the surface state: (i) spin-orbit induced spin mixing within the bulk wave functions themselves\cite{yaz10} and, (ii), an intrinsic topological spin entanglement arising from the superposition of bulk band edge states that comprise the topological surface state. Secondly, we find that the microscopically derived surface state wave function has, \emph{for the same crystal facet}, a highly non-universal richness of spin structure: not only a helical spin texture -- the ``standard result'' -- but also hyperbolic and linear spin textures. Which of these is realized depends sensitively on the microscopic physics of the material: the balance of spin-orbit and crystal field effects in the bulk, and the band bending at the crystal-vacuum interface.

\emph{Band structure model of the IV-VI semiconductors}: An analytically tractable yet fully atomistic theory of the IV-VI semiconductors has been developed by Pankratov and co-workers\cite{vol78,vol83,vol85,pan87,pan90}, and this will form the basis of our discussion of these materials. The bulk spectrum possesses a direct band gap at the $L$ points of the face centred cubic Brillouin zone (BZ), at which there exist Kramers degenerate band edge states that are of pure group IV or group VI character, and for which parity is a good quantum number. By convention positive (negative) parity labels the group VI (group IV) species. These band edge states are expressed in terms of the standard linear combinations of the Bloch functions derived from atomic $p$-orbitals $\Phi_0^{\pm\uparrow\downarrow}=\Phi_{z}^{\pm\uparrow\downarrow}\), $\Phi_\pm^{\pm\uparrow\downarrow}=\mp\left(\Phi_{x}^{\pm\uparrow\downarrow}\pm i \Phi_{y}^{\pm\uparrow\downarrow}\right)/\sqrt{2}$ as

\begin{align}
\Phi_2^- &=-\sin\frac{\Theta^-}{2}\Phi_+^{-\downarrow}+\cos\frac{\Theta^-}{2}\Phi_0^{-\uparrow} \label{A}\\
K\Phi_2^- &=-\sin\frac{\Theta^-}{2}\Phi_-^{-\uparrow}+\cos\frac{\Theta^-}{2}\Phi_0^{-\downarrow} \label{B} \\
\Phi_1^+ &=\ \ \cos\frac{\Theta^+}{2}\Phi_+^{+\downarrow}+\sin\frac{\Theta^+}{2}\Phi_0^{+\uparrow} \label{C}\\
K\Phi_1^+ &=\ \ \cos\frac{\Theta^+}{2}\Phi_-^{+\uparrow}+\sin\frac{\Theta^+}{2}\Phi_0^{+\downarrow}
\label{D}
\end{align}
where \(K\) is the Kramers operator, the superscript \(\pm\) labels parity, and the superscript \(\uparrow\downarrow\) the z-component of the spin which is quantized along the (111)-axis, a natural coordinate system for electronic structure at the $L$ point. The spin ad-mixture of these gap edge wave functions is controlled by two material dependent \emph{spin mixing parameters} \(\Theta^\pm\), that depend on the ratio of the crystal field mixing of the $p$-orbitals $w^\pm$
to the spin-orbit coupling strength for each atomic species $\hbar \lambda^\pm$:

\begin{align}
&\tan\Theta^\pm=-\frac{2\sqrt{2}}{1+3(w^\pm/\hbar \lambda^\pm)}.
\label{SM}
\end{align}
In the basis of the four bulk band edge states given by Eq.~(\ref{A}-\ref{D}) the low energy electronic structure of the bulk material is described by a Dirac Hamiltonian

\begin{align}
H=\begin{pmatrix}
\Delta_0&\hbar\ \boldsymbol{\tau}.V\mathbf{k}\\
\hbar\ \boldsymbol{\tau}.V\mathbf{k}&-\Delta_0
\end{pmatrix}
\label{H_D}
\end{align}
in which all objects are referred to the local $L$-point coordinate system: \(\boldsymbol{\tau}\) is the vector of Pauli matrices $(\tau_x,\tau_y,\tau_z)$ and $V$ a velocity matrix $V = \text{Diag}(v_\perp,v_\perp,v_\parallel)$ where \(v_\parallel\) (\(v_\perp\)) stands for the velocity parallel (perpendicular) to the (111)-direction. The bulk band gap $2\Delta_0$ is defined as the difference in energy between the $L_{6\pm}$ band edge states: $2\Delta_0 = \epsilon_{L_{6-}} - \epsilon_{L_{6+}}$.

\emph{The topological surface states}:
There are four non-equivalent $L$-points in the bulk band structure, each of which generates its own Dirac cone surface state. To focus exclusively on the bulk-boundary correspondence we consider a surface for which these Dirac cones do not interact; amongst the high symmetry facets this is realized by the (111) surface for which the bulk $L$ points project to the $\Gamma$ and to the three inequivalent $M$ points of the hexagonal surface BZ\cite{tan13}. 
The topological boundary condition\cite{vol85,zha12} then allows us to model the bulk-boundary problem simply by replacing  the constant $\Delta_0$ in Eq.~\eqref{H_D} by a $z$ dependent \emph{gap function} \(\Delta(z)=\Delta_0f(z)\), where $z$ is the direction normal to the surface and $f(z\to-\infty)\to-\infty$ (the vacuum side) and $f(z\to\infty)\to1$ (the material side). The surface is defined by the fact that $\Delta(z\to -\infty) = \infty$ on the vacuum side while $\Delta(z\to \infty) = \Delta_0$, the bulk band gap, on the material side. A topologically non-trivial interface requires band inversion on the material side with respect to the vacuum side\cite{zha12}. We therefore have $\Delta(-\infty)\Delta(\infty) <0$ implying $\Delta_0 < 0$; for the IV-VI materials this is satisfied by Pb$_{1-x}$Sn$_x$Te\cite{tan13a}, and Pb$_{1-x}$Sn$_x$Se\cite{pol14} (for large enough $x$). Finally, we introduce an energy shift function \(\varphi(z)=\varphi_0f(z)\) in order to model band bending at the surface. The choice of the same scaling function for $\Delta (z)$ and $\phi (z)$ does not affect generality since, as we will show, $f(z)$ does not enter the in-plane physics of the DW surface state.

\begin{figure*}[t!]
        \includegraphics[width=\linewidth]{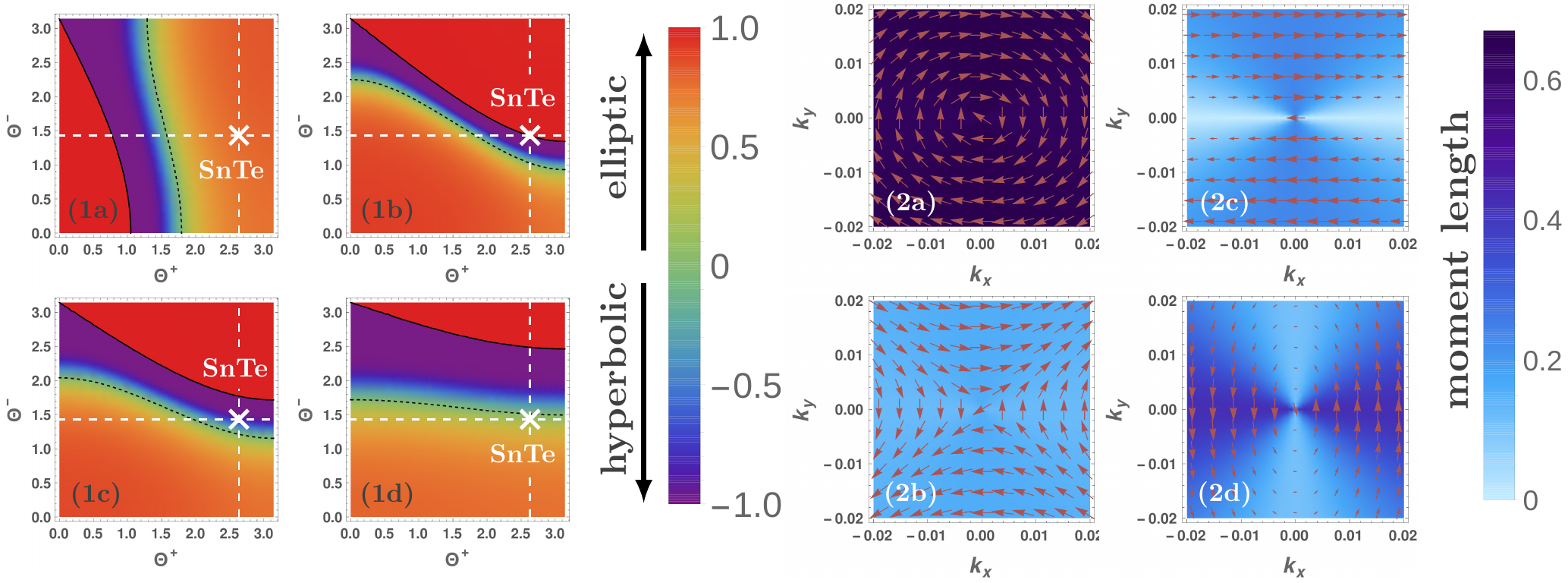}
  	\caption{\emph{Dependence of the topological spin texture for the (111) surface $M$-point Dirac cone on the the microscopic physics of the IV-VI semiconductor}. The rich diversity of spin textures (see Eq.~\ref{spin-pol}) may be characterized by a ``texture number'' $\eta = \tanh b/a$, positive for helical and negative for hyperbolic textures. The left hand panels display a density plot of $\eta$ versus the spin mixing angles $\Theta^\pm$, which encode the balance of spin-obit interaction and crystal field in the bulk -- see Eq.~\eqref{SM}. Each panel represents a different value the surface band bending parameter: $\varphi_0/\Delta_0 = 0.60$, $-0.24$, $-0.40$, $-0.80$ (panels (1a) to (1d) respectively). The panels on the right show the spin texture calculated using $\Theta^\pm$ that describe the IV-VI semiconductor SnTe\cite{pan87,pan90}, and are shown for each of the $\varphi_0/\Delta_0$ presented in (1a-1d). The background density plot represents the magnitude of the spin and the arrows the direction. The spin texture is in each case plotted for the electron cone (and would change sign on the hole cone). Note that the winding number of the texture changes from -1 (2a) to +1 (2b) as the band bending changes from downwards to upwards.}
	\label{eta}
\end{figure*}

To implement this within Eq.~\ref{H_D} for an arbitrary crystal facet requires a coordinate transformation such that the $k_z$ axis coincides with the surface normal. This is achieved in two steps: (i) a rotation $R_z(\alpha)$ about the (111) axis followed (ii) by a rotation $R_y(\beta)$ about the new $k_y$ axis. In this way we obtain the following differential equation in $z$:

\begin{align}
\begin{pmatrix}
\Delta(z)& \lambda\\
\lambda&-\Delta(z)
\end{pmatrix}
\psi=\left(\epsilon-\varphi(z)\right)\psi,
\label{EVP_shift}
\end{align}
where \(\lambda=A - i B \partial_z\) with \(A=\hbar \left(\boldsymbol{\tau}.RVR^\dagger \mathbf{k}_\perp\right)\) and \(B=\hbar \left(\boldsymbol{\tau}.RVR^\dagger \mathbf{e}_z\right)\), with $R$ the composite rotation operator and where we introduced the standard notation \(\mathbf{k}_\perp=k_x\mathbf{e}_x+k_y\mathbf{e}_y\). This eigenvalue problem may be solved by a similar strategy to that deployed in Ref.~\onlinecite{pan87,pan90}: Eq.~(\ref{EVP_shift}) is squared and a unitary operator applied that yields two decoupled spinor equations. The required transformation is \(\zeta=S\psi\) with

\begin{align}
S=\sqrt{\frac{\Delta_0}{2\left(\varphi_0^2-\Delta_0^2\right)}}
\begin{pmatrix}
\sqrt{\varphi_0+\Delta_0}&\sqrt{\varphi_0-\Delta_0}\\
-\sqrt{\varphi_0+\Delta_0}&\sqrt{\varphi-\Delta_0}
\end{pmatrix}\otimes D
\label{S-xi-theta}
\end{align}
where $D=e^{i\kappa z}\left[a_+\left(1-i\tau_z\right)+ia_-\left(\tau_y-\tau_x\right)\right]/2$
with \(a_\pm(\beta)=\sqrt{1\pm(v_\perp\sin^2\beta+v_\parallel\cos^2\beta)/v_1}\) and \(v_1=\sqrt{v_\perp^2\sin^2\beta+v_\parallel^2\cos^2\beta}\). In the solution to Eq.~\eqref{EVP_shift} the parameter $\kappa \in \mathbb{R}$ will turn out to encode a mixing of real momentum into the exponential decay envelope of the surface state, but for now represents a free parameter. This procedure results in two separate equations for the two spinors \(\zeta_\pm\) in $\zeta=(\zeta_+,\zeta_-)$:

\begin{align}
\notag &\big[B^2(\kappa-i\partial_z)^2+W(z)^2\mp \hbar v_1 \sigma_z W(z)\\
&+\{A,B\}(\kappa-i\partial_z)\big]\zeta_\pm=\left[\frac{\epsilon^2\varphi_0^2}{\Delta_0^2-\varphi_0^2}-A^2+\epsilon^2\right]\zeta_\pm
\label{SSE_AB}
\end{align}
where \(W(z)=\sqrt{\Delta_0^2-\varphi_0^2}\left[f(z)+\epsilon\varphi_0/\left(\Delta_0^2-\varphi_0^2\right)\right]\). The linear in $k_z$ terms of Eq.~\ref{SSE_AB} may be removed by the freedom to choose $\kappa$, and in this way the eigenvalue problem can be brought to the familiar equation of Witten's super-symmetric quantum mechanics:

\begin{align}
\notag &\Bigl[-\hbar v_1 \sigma_z \partial_z\pm W(z)\Bigr]\Bigl[+\hbar v_1 \sigma_z \partial_z \pm W(z)\Bigr]\zeta_\pm\\
&=\left[\epsilon^2-A^2+\frac{\{A,B\}^2}{4B^2}+\frac{\epsilon^2\varphi_0^2}{\Delta_0^2-\varphi_0^2}\right]\zeta_\pm
\label{SSE_W1}
\end{align}
The ground state of the positive semi-definite operator on the left hand side follows from the equation $\left[\hbar v_1 \sigma_z\partial_z \pm W(z)\right]\zeta_\pm=0$, for which there are only two linearly independent normalizable solutions for the decay envelope of the surface state: $\zeta_{-s} = c_- \begin{pmatrix} 0, 1 \end{pmatrix}^T g(z)$ and $\zeta_{s} = c_+ \begin{pmatrix} 1, 0 \end{pmatrix}^T g(z)$ where $s = 2H(\varphi_0)-1$, with $H(x)$ the Heaviside function. Substitution of either of these yields a differential equation for the surface state decay function given by $\left[\hbar v_1 \partial_z + W(z)\right]g(z)=0$, and yields the solution $g(z)=N\exp\left[{-s\frac{1}{\hbar v_1}\int_0^z dz'\ W(z')}\right]$
where we have defined the normalization $N=\left(\int_{-\infty}^\infty dz\ \mathrm{exp}\left[-2s/(\hbar v_1)\int_{0}^z dz'\ W(z')\right]\right)^{-1/2}$. The requirement for the surface state to be normalizable on \emph{both} the material and vacuum sides demands that $W(z)$ changes sign asymptotically. The presence of the energy shift function, however, complicates this ``natural'' topological boundary condition requiring in addition that: (i) the band bending at the surface must be less than the gap function i.e. $|\varphi_0|<|\Delta_0|$ and (ii) $\epsilon_{-s} < \epsilon <\epsilon_{s}$ with $\epsilon_{s}=-\varphi(-s\infty)\left(\Delta_0^2/\varphi_0^2-1\right)$, i.e. the Dirac cone now exists in a semi-infinite energy range, merging with the valence (conduction) bulk bands for $s=+1$ ($s=-1$).

For the full surface state (and not just its $z$ dependent envelope) we evidently require the coefficients of the spinor solutions $\zeta_\pm$. The decoupled problem cannot determine these coefficients and we must therefore substitute the wave function $\psi = S^{-1}(\zeta_+,\zeta_-)^T$ back into Eq.~(\ref{EVP_shift}). This results in a system of 4 equations of which only two are independent, and may be written as

\begin{align}
\hbar\gamma 
\begin{pmatrix}
0&v_xk_x-iv_\perp k_y\\
v_x k_x+iv_\perp k_y&0
\end{pmatrix} \begin{pmatrix} c_+ \\ c_- \end{pmatrix}
= \epsilon \begin{pmatrix} c_+ \\ c_- \end{pmatrix}
\label{DD}
\end{align}
where the basis for this equation is obtained from the transformation from pseudospin back to atomic variables. In this way we find the basis functions linking the pseudospin and atomic variables to be given by the Kramers conjugate pair of functions $X$ and $K X$ where:
\begin{equation}
X = s F_- G_+ \Phi_2^- + s F_- G_- K\Phi_2^- + F_+ G_+ \Phi_1^+ + F_+ G_- K\Phi_1^+
\label{X}
\end{equation}
and where $F_\pm = 1/2\sqrt{1\pm\varphi_0/\Delta_0}e^{\mp i\pi/4}$ depends only on the surface band bending, and $G_{\pm} =\left(a_\pm(\beta)\cos\beta/2 \mp  a_\mp(\beta)\sin\beta/2\right) e^{\mp i \alpha/2}$ only on the angles $\alpha$ and $\beta$ describing misorientation between the principal axis of the $L$ point Fermi pocket and the surface normal.

Eq.~\eqref{DD} evidently describes a DW surface state, and thus for an arbitrary facet of a IV-VI semiconductor we find that the spectrum is an anisotropic Dirac cone

\begin{align}
\epsilon_{\mathbf{k}_\perp l}=l\hbar \gamma\sqrt{v_x^2k_x^2+v_\perp^2k_y^2}
\label{spec}
\end{align}
with $v_x = (v_\parallel v_\perp)/v_1$ and \(\gamma=\sqrt{1-\varphi_0^2/\Delta_0^2}\). The anisotropy is controlled by the ratio of the two velocities \(v_x/v_\perp\) which, through \(v_x\), depends on the surface orientation. In pseudospin space the solution of Eq.~\ref{DD} is

\begin{align}
\begin{pmatrix} c_+ \\ c_- \end{pmatrix} =\frac{1}{\sqrt{2}}
\begin{pmatrix}
&1\\
&le^{i\phi}
\end{pmatrix}
\label{chi}
\end{align}
with \(\phi=\tan^{-1}[v_\perp k_y/(v_x k_x)]\). The microscopic surface state wavefunction is, however, not the pseudospin solution of the DW equation but this solution expressed in terms of the basis functions $X$ and $KX$ of the DW equation:

\begin{equation}
\Psi_{\mathbf{k}_\perp l} = (c_+ X + c_- KX) e^{i\mathbf{k}_\perp.\mathbf{r}_\perp}e^{i\kappa z}g(z)
\label{yolo}
\end{equation}
where the effective pseudospin degree of freedom has given way to the microscopic physical spin encoded in the bulk gap edge wavefunction through Eq.~\eqref{X}. These expressions, Eq.~\eqref{X} and Eq.~\eqref{yolo}, are the central results of this paper: they provide an explicit connection between the topological surface state and the microscopic  physics of the semiconductor. The spin structure of the surface state will, therefore, be determined by the superposition of gap edge states, and thus be governed by Eq.~\eqref{X}. Such spin mixing will involve both that intrinsic to the bulk states (spin-orbit coupling driven) as well that from the superposition of bulk states (spin mixing from the topological boundary condition). This relation is, one should note, applicable to any bulk Hamiltonian of the form given by Eq.~\eqref{H_D} (although the gap edge basis functions will evidently change from material to material), and so represents a rather general relation between bulk and surface states in topological insulators, applicable, for instance, to the Bi$_2$Se$_3$ TI class (if $p^2$ corrections are ignored)\cite{liu10,zha12}.

\emph{The spin polarization}: The expectation of the physical spin operator is now easily calculated from Eq.~\eqref{yolo}:

\begin{equation}
\langle\Psi_{\mathbf{k}_\perp l}\mid \boldsymbol{\sigma}\mid\Psi_{\mathbf{k}_\perp l}\rangle=
l
\begin{pmatrix}
  -a \cos\phi\\
 b \sin\phi\\
m_z \sin\phi
\end{pmatrix}
\label{spin-pol}
\end{equation}
where $l=\pm 1$ for the electron and hole cone respectively, \(a=\rho_1 (v_x/v_\parallel)\sin^2\beta+\rho_2 (v_x/v_\perp)\cos^2\beta\), \(b=\rho_2\), and $m_z =\sin(2\beta)\left[(v_x/v_\parallel)\rho_1-(v_x/v_\perp)\rho_2\right]/2$. In these expressions we have defined the constants $\rho_1 = \frac{1}{2}(1+\frac{\varphi_0}{\Delta_0})\cos\Theta^+ + \frac{1}{2}(1-\frac{\varphi_0}{\Delta_0}) \cos\Theta^-$ and $\rho_2 = -\frac{1}{2}(1+\frac{\varphi_0}{\Delta_0})\sin^2\frac{\Theta^+}{2} + \frac{1}{2}(1-\frac{\varphi_0}{\Delta_0})\cos^2\frac{\Theta^-}{2}$. From Eq.~\ref{spin-pol} it is seen that the spin polarization consists of an in-plane component given by the tangent field to the conic section form \(k_x^2/a^2+\mathrm{sign}(ab)k_y^2/b^2=1\), and an out-of-plane component governed by $m_z$. While formally similar results have been presented before, the novel feature of Eq.~\ref{spin-pol} lies in the fact that the spin texture is now determined by all the microscopic variables of the IV-VI semiconductor: the anisotropy of the $L$-point Fermi surface ($v_\parallel$ and $v_\perp$), the balance of spin-orbit and crystal field interactions ($\Theta^\pm$) in the bulk, the angle between the crystal facet and the principle axis of the $L$-point Fermi pocket ($\beta$), and the band bending at the surface ($\varphi_0/\Delta_0$).

We will first consider the spin texture of the $\Gamma$-point Dirac cone on the (111) facet, $\beta = 0$. The spin polarization is then in-plane ($m_z = 0$) with polarization parameters $a = b = \rho_2$, i.e. a circular spin texture of winding number $-\text{sign}(\rho_2)$. Inserting the known values of $\Theta^\pm$ for SnTe\cite{pan87} in the formula for $\rho_2$, yields, for downward band bending ($\varphi_0 > 0$) as seen in experiment\cite{tas14}, a winding number of -1 (+1) for the conduction (valence) electrons and a polarization of $\approx 0.8\mu_B$ -- in agreement with ab-initio calculations and experiment\cite{shi14}. Note that the band bending significantly influences the spin texture: for $\varphi_0/\Delta_0 = -(\sin^2\frac{\Theta^+}{2}-\cos^2\frac{\Theta^-}{2})/(\sin^2\frac{\Theta^+}{2}+\cos^2\frac{\Theta^-}{2})$ we even have $\rho_2=0$, i.e. the topological polarization of the surface state vanishes. The agreement with experiment for the experimental band bending thus represents a rigorous test of the theory.

Having thus validated our model, we now consider the $M$-point Dirac cones. These have $\cos\beta=1/3$ which results, as we now show, in a profoundly rich and highly tunable spin texture. Firstly, a non-zero value of $\beta$ implies that all possible sign combinations of $a$ and $b$ can be realized which, as the spin texture is the tangent field to \(k_x^2/a^2+\mathrm{sign}(ab)k_y^2/b^2=1\), implies that circular, elliptical, hyperbolic, as well as linear spin textures are possible. The left hand panels of Fig.~\ref{eta} display the qualitative form of this spin texture as a function of the spin mixing parameters $\Theta^\pm$, plotted for four different values of the band bending (see figure caption for details). Evidently both of these microscopic variables profoundly influence the qualitative form of the spin-momentum locking in the topological surface state.

\begin{figure}[h]
        \includegraphics[width=0.98\linewidth]{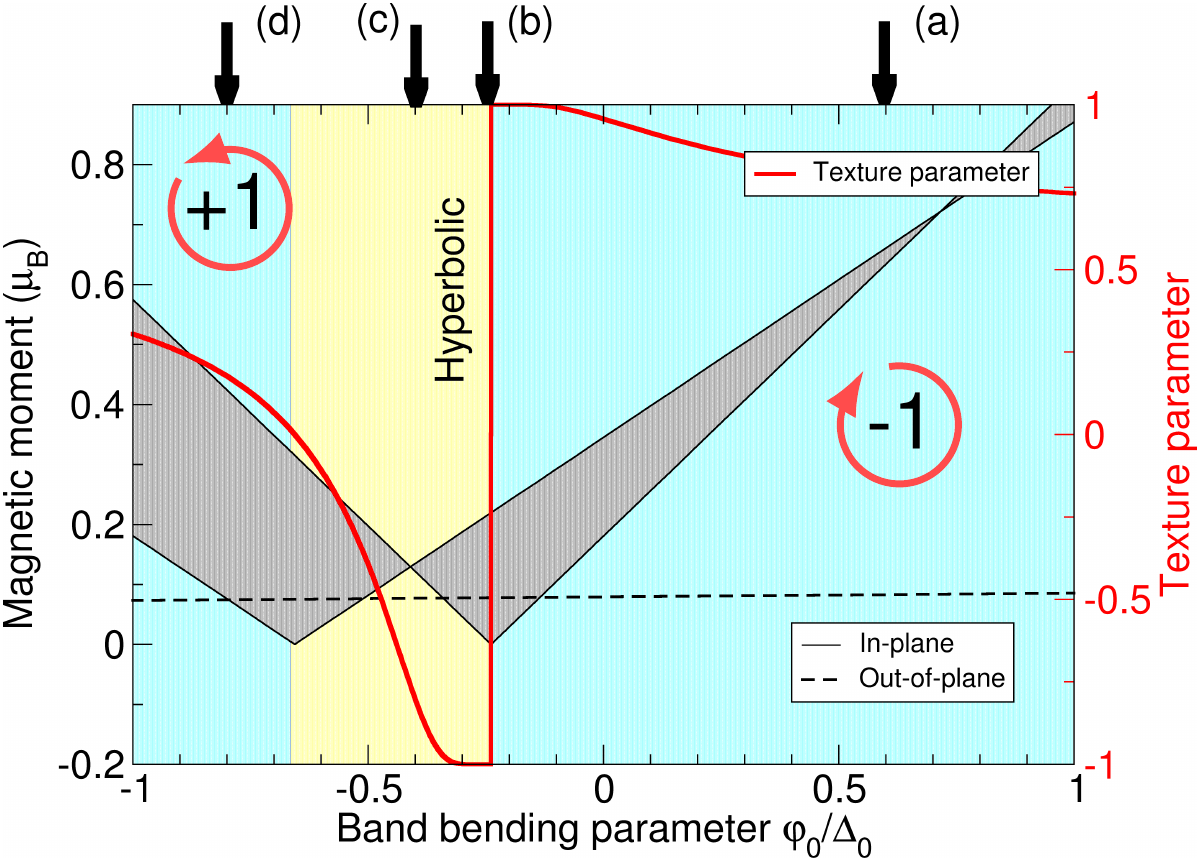}
  	\caption{Surface polarization and spin texture parameter as a function of band bending for the $M$-point cone of the (111) surface of SnTe. The shaded region represents the range of in-plane polarization, while the dashed dark line the maximum magnitude of the out-of-plane component. The light (red) line shows the spin texture parameter which is positive for a helical texture, and negative for a hyperbolic texture (the highlighted region). Note that the winding number of the spin texture changes from -1 to +1 on going though the hyperbolic region. Arrows (a-d) indicate the spin textures plotted in panels (2a-2d) of Fig.~1.}
	\label{field}
\end{figure}

The spin texture for SnTe 
is shown in the four right hand panels for the same four values of the band bending. Remarkably we see that simply by tuning the band bending from downward to upward, the spin texture evolves from helical with winding number -1 (2a), through linear (2a) and hyperbolic (2c) textures, to finally a helical texture of winding number +1 (2d). The full evolution of the spin texture parameter $\eta = \tanh b/a$ and spin polarization as a function of band bending is shown in Fig.~\ref{field}. The spin polarization has a non-trivial out-of-plane component (see Fig.~\ref{field}) which, interestingly, is in agreement with a recent ab-initio calculation for the $M$ point\cite{shi14}. While the magnitude of this is approximately constant ($\approx 0.1\mu_B$), the in-plane moment is significantly larger for downward ($\varphi_0 > 0$) as opposed to upward ($\varphi_0 < 0$) band bending. In short, the spin-momentum locking is highly non-universal and, furthermore, highly tunable, for example by application of a top gate or surface doping, both of which would alter the surface band bending.

\emph{Conclusions}: On the basis of an exact relation between surface state and bulk gap edge wave functions, valid for a wide class of TI's, we elucidate a precise relation between pseudospin and physical spin. The surface state spin structure is found to emerge both from spin mixing within bulk wave functions (bulk spin-orbit driven), as well as from the superposition of these wave functions in the surface state (TI boundary condition driven), and highly sensitive to band bending in the surface region). Thus while the surface Dirac-Weyl spectrum is universal, the surface Dirac-Weyl wave function is highly non-universal, and exhibits significant dependence on all the microscopic variables of the material. The physics of the TI surface state -- impurity scattering, gap opening, as well as direct indicators of the spin texture such as the RKKY interaction -- is thus expected to exhibit a sensitive dependence on bulk microscopic physics and a tunable sensitivity to surface physics.

\bibliography{SnTe-spin-pol-paper}

\end{document}